\documentclass[12pt]{JHEP3}
\usepackage{multicol}
\usepackage[dvips]{epsfig,graphics}
\usepackage{graphicx}
\usepackage{cite}

 \newlength{\wth}
\title{MSSM dark matter measurements at the LHC without squarks and sleptons}
\author{M. J. White and F. Feroz\\
Cavendish Laboratory. Madingley Road. Cambridge CB3 0HE, UK}
\abstract{We examine the case of neutralino dark matter in the focus point region of the MSSM, in which the scalar sparticles are too heavy to be produced at the LHC. Whilst it has been previously asserted that the LHC alone would fail to constrain the properties of the lightest neutralino for such a scenario, we find that one can obtain good predictions of astrophysical quantities such as the relic density, annihilation cross-section and direct search cross-sections by using the shape of the dilepton invariant mass spectrum to constrain neutralino mixing. We demonstrate our technique using a Bayesian analysis of the 24 parameter MSSM model space, and in the process introduce a novel way of improving the LHC results even without assumptions on which new sparticles are responsible for the kinematic features in the dilepton invariant mass distribution.}
\keywords{Beyond the Standard Model, MSSM, Dark Matter}
\preprint{Cavendish-HEP-2010-02}
\begin{document}
\section{Introduction}
The existence of a large amount of non-baryonic `dark' matter (DM) in the universe is now relatively uncontroversial, and the precise nature of the matter is currently the source of much speculation. The WMAP observations of the cosmic microwave background have provided a precise measurement of the dark matter relic density~\cite{Komatsu:2008hk}, and also strongly support cold dark matter in the form of weakly interacting massive particles (WIMPs). Since the Standard Model (SM) of particle physics does not contain any particles that fit this description, searches for dark matter are intimately connected with searches for new physics.

The Large Hadron Collider (LHC) at CERN, Geneva, is designed to probe the TeV scale for evidence of new fundamental theories, and can be expected to produce dark matter particles if a WIMP exists with an accessible mass. At the same time, direct search experiments continue to look for interactions between Earth-based targets and passing WIMPs, indirect search experiments look for evidence of WIMP annihilation in space, and the cosmic microwave background is being mapped with ever greater precision. If it is genuinely true that a single WIMP candidate explains dark matter, one should obtain a consistent set of observations from these experiments, but checking this consistency is hampered by large uncertainties in the relevant astrophysics. For example, if no annihilation signal is observed, how do we know if this is because the annihilation cross-section lies below the reach of our current apparatus or whether we simply looked in a region where there was too little dark matter to give a strong flux? If a direct search experiment fails to see a signal, how do we untangle the interaction cross-section from the uncertainties in the local density and velocity of the dark matter that affect the number of observed events? 

From a particle physics perspective, the only answer is to try and measure the microscopic properties of the WIMP as precisely as possible in accelerator experiments in order to make accurate predictions of the astrophysical data. These measurements can then either be used with halo models to predict annihilation fluxes and direct search cross-sections, or, in the exciting case of a \emph{positive} result in one or more astrophysical experiments, to directly constrain the halo distributions themselves. Comparisons with the relic density could give important clues on the proportion of dark matter composed of the WIMP observed at the LHC or could help validate the standard picture of  Big Bang cosmology. 

This paper examines dark matter arising from the Minimal Supersymmetric Standard Model (MSSM) in the popular case that the WIMP is the lightest neutralino of the model. Previous studies~\cite{baltz-2006-74,nojiri-2006-0603} have indicated that, under favourable conditions, one can measure the WIMP mass and couplings rather well at the LHC but, in the case that the scalars are too heavy to be produced, one might get the mass but certainly not the couplings. A classic example is the `focus point' region of the mSUGRA parameter space~\cite{Feng:2000gh}, in which the scalar masses may be high while the model still obeys all experimental constraints, including the WMAP constraint on the relic density. A linear collider could resolve these issues but is still many years away and, while it would be possible to use astrophysical data in combination with the LHC data, this relies on assuming the model which we are trying to test. Hence there is a clear need to improve DM predictions at the LHC if we are to make proper use of the rich astrophysical data sets expected over the next ten years.

We revisit the focus point SUSY case and demonstrate that the LHC can do significantly better than previously advertised, by using the shape of the dilepton invariant mass spectrum to constrain neutralino mixing. We also find that the improvements are still obtained even if one assumes that the sparticles producing kinematic endpoints at the LHC have not been correctly identified. In analysing our benchmark point we quote a study from members of the ATLAS collaboration~\cite{:2008zzm}, though our results would apply equally well to CMS~\cite{:2008zzk}. We then use the anticipated ATLAS data in a Bayesian analysis of the 24 parameter MSSM space, and hence although we use an mSUGRA benchmark point, our analysis is not restricted to the mSUGRA parameter set. Moreover, it should be possible to apply our approach to other SUSY models with heavy scalars.

The paper is structured as follows. Section~\ref{Background} briefly reviews the relevant background, defines the benchmark point used in this study and introduces the {\sc MultiNest} sampling algorithm~\cite{2009MNRAS.398.1601F} used to obtain our results. In Section~\ref{Results} we describe how to use the shape information to improve the results, and present the posteriors obtained using the `conventional' and `new' approaches. We also discuss potential pitfalls. We present conclusions in section~\ref{Conclusions}.
\section{Background}\label{Background}
\subsection{Neutralino dark matter in the  focus point region}
The Minimal Supersymmetric Standard Model (MSSM) describing broken SUSY permits 108 parameters and, although a large number of these must be small to suppress flavour changing processes, at least 24 parameters are normally considered necessary to investigate LHC phenomenology. Although SUSY provides several WIMP candidates (including the sneutrino and gravitino), the most popular is the lightest neutralino, which is an admixture of the superpartners of the standard model gauge bosons, with components set by the following mixing matrix:
\begin{eqnarray}
{\cal M}=\left(\begin{array}{cccc}
  M_1       &      0          &  -m_Z \cos\beta s_W  & m_Z \sin\beta s_W \\[2mm]
   0        &     M_2         &   m_Z \cos\beta c_W  & -m_Z \sin\beta c_W\\[2mm]
-m_Z \cos\beta s_W & m_Z \cos\beta c_W &       0       &     -\mu        \\[2mm]
 m_Z \sin\beta s_W &-m_Z \sin\beta c_W &     -\mu      &       0
                  \end{array}\right)\
\label{eq:massmatrix}
\end{eqnarray}
where $M_1$ and $M_2$ are the U(1) and SU(2) gaugino masses, $\mu$ is the Higgsino mass parameter, tan$\beta$ is the ratio of the vacuum
expectation values of the two Higgs doublets and the other parameters are all from the standard model. Measuring the couplings of a neutralino WIMP essentially comes down to trying to constrain the components of this matrix which is not generically possible at the LHC.

Many phenomenological studies have been performed using simpler frameworks such as the minimal supergravity model (mSUGRA)~\cite{mSUGRA} where SUSY breaking, which is assumed to occur in a hidden sector, is communicated to the observable sector via gravitational interactions. The mSUGRA model unifies various GUT scale parameters, obtaining the following parameter set: the scalar mass $m_0$, the gaugino mass $m_{1/2}$, the trilinear coupling $A_0$, the ratio of Higgs expectation values tan$\beta$, and the sign of the SUSY Higgs mass parameter $\mu$. Assuming the lightest neutralino is the correct WIMP candidate, one finds that, in most of the mSUGRA parameter space, too many neutralinos would survive after the Big Bang, and hence the only allowed regions must have some kind of annihilation mechanism to bring the density down to within the limits set by astrophysical observation. There are four main mechanisms, each of which dominates in a particular region of the parameter space~\cite{baltz-2006-74}:
\begin{enumerate}
\item Slepton exchange. This is suppressed unless the slepton masses
  are lighter than approximately 200 GeV and gives rise to the `bulk' region at low values of the mass parameters.
\item Co-annihilation with light sleptons. This occurs when there are
  suitable mass degeneracies in the sparticle spectrum, giving rise to a suitably named `co-annihilation region'.
\item Annihilation to third-generation fermions. This is enhanced when
  the heavy Higgs boson $A$ is almost twice as massive as the LSP, giving rise to a `funnel region' for models with reasonably large tan$\beta$.
\item Annihilation to vector bosons. This can occur if the neutralino
  LSP acquires a significant wino or higgsino component, which happens in the so-called `focus point' region at high $m_0$.
\end{enumerate}
One must remember that the tight level of constraint provided by the WMAP data is only a feature of models with a small number of parameters. A recent review of mSUGRA results that investigates the effect of adding new parameters can be found in reference~\cite{Baer:2008ih}, while the results of a 24 parameter Bayesian analysis incorporating all current data can be found in~\cite{AbdusSalam:2009qd}.

\FIGURE[ht]{\epsfig{file=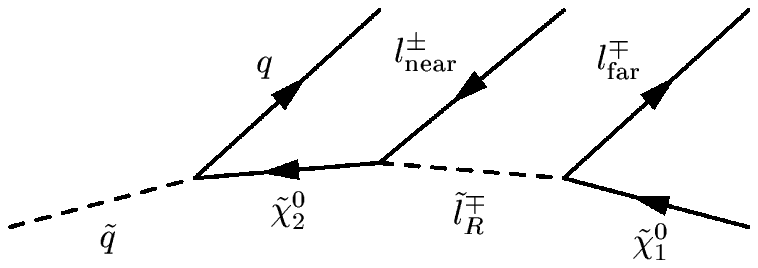,width=4in}\caption{A popular cascade decay chain \label{fig:decay}used to measure sparticle masses at the LHC. The invariant mass distributions of the decay products will exhibit kinematic endpoints whose position is a function of the four sparticle masses involved in the cascade.}}

The high value of $m_0$ in the focus point region puts the squark and slepton masses in the multi-TeV range and thus out of direct reach of the LHC, and this forms the basis of why the LHC is expected to struggle to measure the WIMP couplings in this region. The main method of constraining the parameters of the mixing matrix is to look for cascade decay processes featuring squarks and sleptons (see figure~\ref{fig:decay}) and use kinematic endpoints in the invariant mass distributions of the SM decay products to measure mass differences between the sparticles. If a large amount of the sparticle mass spectrum can be measured (for example in the bulk or coannihilation regions), one can tell that the LSP is almost purely bino and reconstruct the relic density to a precision of around 10\%~\cite{nojiri-2006-0603}, dependant on assumptions about the LHC reach in the Higgs sector. This remains true in a 24 parameter MSSM analysis, and thus is not limited to an unreasonably restrictive mSUGRA analysis. Mass differences can still be measured in the focus point region using cascade decays headed by gluinos, but the lack of squarks and sleptons will clearly reduce the number of decays that we can observe, and it was previously argued that this leaves the dominant component of the neutralino ambiguous when one attempts to fit the SUSY parameters~\cite{baltz-2006-74}, leading to substantial (i.e. orders of magnitude) uncertainties in the gamma ray flux, direct detection cross-section and relic density results from the LHC.

Nevertheless, the focus point is attractive for a number of reasons. From a DM perspective, it is interesting to note that the higgsino component of the neutralino WIMP implies large annihilation via $W^+W^-$ and $ZZ$ bosons, and a large number of gamma rays in the final state, thus improving the prospects for finding dark matter using the forthcoming generation of gamma ray telescopes. Indeed, a previous study comparing the search reach of the LHC and H.E.S.S. II revealed a region of overlap in the focus point region that would give clear signals in both experiments~\cite{moulin-2008-77}. If a focus point type scenario were chosen by Nature, we therefore have the potential to learn much about dark matter. Furthermore, the focus point region is actually favoured by constrained MSSM fits to particle physics and cosmological data if one assumes `natural priors'~\cite{Allanach:2008iq}.

%\begin{figure}
%\centerline{
%\includegraphics[width=2.4in]{mode1.eps}
%}
%\caption{A popular cascade decay chain used to measure sparticle masses at the LHC. The invariant mass distributions of the decay products will exhibit kinematic endpoints whose position is a function of the four sparticle masses involved in the cascade.
%\label{fig:decay}}
%\end{figure}
\subsection{Benchmark point}
\TABLE[h!]{\begin{tabular}{|c|c|}
\hline
Particle&Mass/GeV\\
\hline
$\tilde{\chi}_1^0$&106\\
$\tilde{\chi}_2^0$&165\\
$\tilde{\chi}_3^0$&191\\
$\tilde{g}$&850\\
$\tilde{b}_1$&2710\\
$\tilde{b}_2$&3241\\
$\tilde{t}_1$&1977\\
$\tilde{t}_2$&2720\\
$\tilde{\tau}_1$&3252\\
$\tilde{\tau}_2$&3268\\
$h$&118.7\\
\hline
\end{tabular}
\caption{Particle masses for the benchmark point described in the text. The squarks and sleptons of the first two generations all have a mass greater than 3 TeV.\label{tablemasses}} 
}
Our study uses a particular benchmark point that gives a typical focus point phenomenology, and a discussion of the generality of our results is deferred to section~\ref{Discussion}. One of the problems of performing studies of the focus point region is that the results of spectrum calculations are highly sensitive to the top Yukawa coupling, and therefore to the physical top mass. Given that one of the leading aims of this paper is to show improvement over previously published results, we have chosen to use the same spectrum generator as that used in~\cite{baltz-2006-74} (\tt ISAJET 7.69\rm~\cite{baer-2003}), with the same focus point region benchmark point (their `LCC2') and the same value of the top mass. While it is true that a more recent and improved version of \tt ISAJET \rm is available, along with a revised measurement of the top mass, the results of this paper should be viewed as comparitive rather than definitive. We thus use the following values of the mSUGRA parameters:
\begin{equation}
m_0=3280 GeV, m_{1/2}=300 GeV, \mbox{tan}\beta = 10, A_0=0, \mu > 0
\end{equation}
with a top mass of 175 GeV. We ran \tt ISASUGRA \rm with these parameters, generating the equivalent 24 parameter MSSM input for \tt ISASUSY\rm. We then used this input to generate the expected values for the observables used in our analysis~\footnote{Note that we do not set the top, bottom and tau masses, Higgs vacuum expectation value and strong coupling constant to the benchmark value as in~\cite{baltz-2006-74}, nor do we scale the Yukawa couplings to more closely match mSUGRA.}. The \tt ISASUSY \rm parameter set gives a slightly different mass spectrum and a relic density $\Omega_{\chi}h^2 = 0.089$ which is a little lower than the WMAP limit but this does not affect the validity of our conclusions. Important masses are given in Table~\ref{tablemasses}, as generated from \tt ISASUSY\rm.

%\begin{table}
%\centerline{
%\begin{tabular}{|c|c|}
%\hline
%Particle&Mass/GeV\\
%\hline
%$\tilde{\chi}_1^0$&106\\
%$\tilde{\chi}_2^0$&165\\
%$\tilde{\chi}_3^0$&191\\
%$\tilde{g}$&850\\
%$\tilde{b}_1$&2710\\
%$\tilde{b}_2$&3241\\
%$\tilde{t}_1$&1977\\
%$\tilde{t}_2$&2720\\
%$\tilde{\tau}_1$&3252\\
%$\tilde{\tau}_2$&3268\\
%$h$&118.7\\
%\hline
%\end{tabular}
%}
%\caption{Particle masses for the benchmark point described in the text. The squarks and sleptons of the first two generations all have a mass greater than 3 TeV.\label{tablemasses}} 

%\end{table}
\subsection{Constraining the MSSM parameter space}
Given a set of LHC observables $\mathbf{O}$ and our set of hypothesised MSSM parameters $\mathbf{P_{MSSM}}$, we will shortly wish to evaluate $p(\mathbf{P_{MSSM}}|\textbf{O})$ and thus to find the region of parameter space consistent with the data. This posterior probability is related to the likelihood $\mathcal{L}(\mathbf{P_{MSSM}})=p(\mathbf{O}|\mathbf{P_{MSSM}})$ by Bayes theorem:
\begin{equation}
p(\mathbf{P_{MSSM}}|\mathbf{O}) = \frac{p(\mathbf{O}|\mathbf{P_{MSSM}})p(\mathbf{P_{MSSM}})}{Z}
\end{equation}
where $p(\mathbf{P_{MSSM}})$ represents our prior knowledge on the distribution of parameters, and the normalization constant $Z$ is the `Bayesian evidence', given by the average of the likelihood over the prior:
\begin{equation}
Z=\int \mathcal{L}(\mathbf{P_{MSSM}}) p(\mathbf{P_{MSSM}})d^{N_{par}}\mathbf{P_{MSSM}}
\label{evidence}
\end{equation}
In our study, we assign points the following likelihood:
\begin{equation}
p(\mathbf{O}|\mathbf{P_{MSSM}})=\prod_i \mbox{exp} \left ( -\frac{(O_i(\mathbf{P_{MSSM}})-o_i)^2}{2\sigma_i^2}\right ) 
\label{likelihood}
\end{equation}
where $O_i(\mathbf{P_{MSSM}})$ is the predicted value of the $i$th observable $o_i$, $\sigma_i$ is the error in $O_i$ and the product runs over all observables.  

In typical SUSY phenomenology problems such as that presented here, the posterior cannot be evaluated analytically and one can instead use a sampling approach. Most popular are Markov Chain Monte Carlo algorithms that draw samples in such a way that the distribution of samples tends towards the unnormalised posterior distribution. This allows one to make inferences on parameter constraints while bypassing the evaluation of the Bayesian evidence. Although this has proven successful in the past in MSSM case studies, the methods (such as the Metropolis algorithm) become far less efficient as the dimension of the parameter space increases, and additionally they struggle in cases with a multimodal posterior where the most likely islands of interest occupy a vanishingly small volume of the parameter space. Sadly, this is the quintessential MSSM posterior, and previous examples have either used many CPU years or have varied the step size as the sampling progresses from a large initial value tuned to find disparate islands to a small value designed to explore the region around each mode. One still has to choose sensible initial conditions for these algorithms, as well as finding sensible step sizes in each parameter if one chooses not to vary them.

A more recent approach is that of `nested sampling' which targets the Bayesian evidence first, subsequently obtaining samples from the posterior as a by product. We here give a brief description and refer the reader to~\cite{2008MNRAS.384..449F, 2009MNRAS.398.1601F} for more details.

\FIGURE{{\includegraphics[width=0.3\columnwidth]{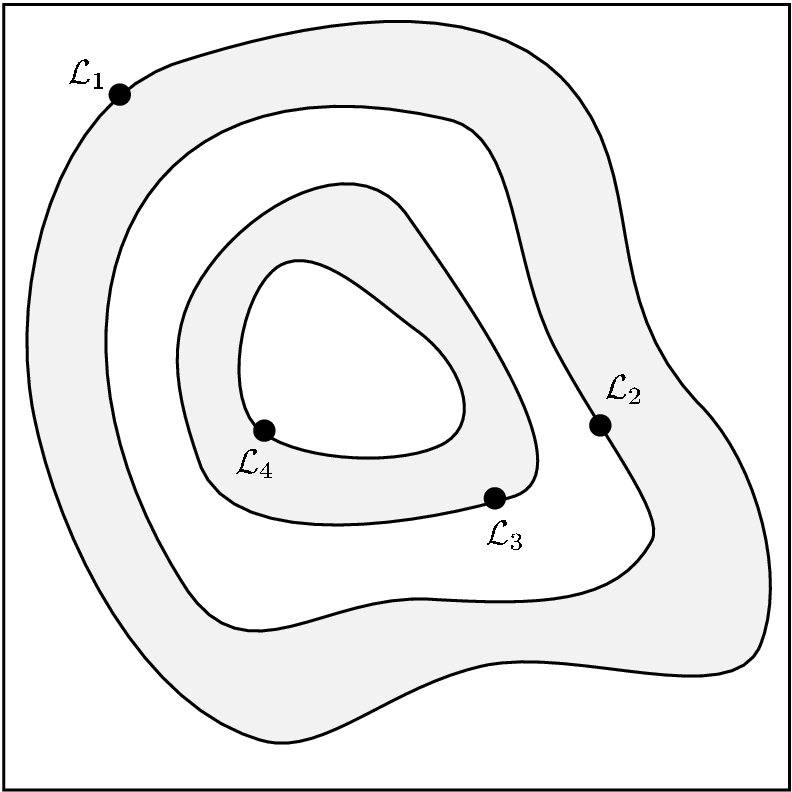}}
\hspace{0.3cm}
{\includegraphics[width=0.3\columnwidth]{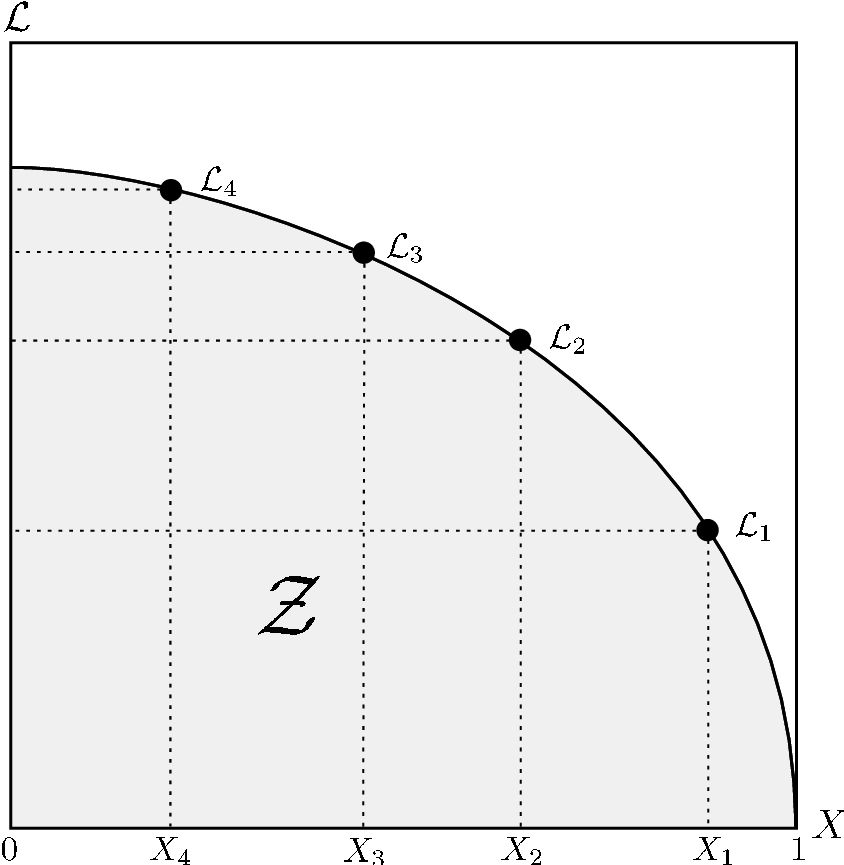}}
\caption{Cartoon illustrating (a) the posterior of a two dimensional problem; 
and (b) the transformed $L(X)$ function where the prior volumes $X_i$ are
associated with each likelihood $L_i$.}
\label{fig:NS}}

Nested sampling~\cite{2004AIPC..735..395S} is a Monte Carlo method that calculates the evidence by
transforming the multi--dimensional evidence integral into a one--dimensional integral that is easy to evaluate numerically. This is accomplished by defining the prior volume $X$ as $dX =
p(\mathbf{P_{MSSM}})d^{N_{par}}\mathbf{P_{MSSM}}$, so that
\begin{equation}
X(\lambda) = \int_{\mathcal{L}\left( \mathbf{P_{MSSM}}\right) > \lambda} p(\mathbf{P_{MSSM}}) d^{N_{par}}\mathbf{P_{MSSM}},
\label{eq:Xdef}
\end{equation}
where the integral extends over the region(s) of parameter space contained within the iso-likelihood contour $\mathcal{L}(\mathbf{P_{MSSM}}) = \lambda$. The evidence integral, Eq.~(\ref{evidence}), can then be
written as
\begin{equation}
\mathcal{Z}=\int_{0}^{1}{\mathcal{L}(X)}dX,
\label{eq:nested}
\end{equation}
where $\mathcal{L}(X)$, the inverse of Eq.~(\ref{eq:Xdef}), is a  monotonically decreasing function of $X$.  Thus, if one can evaluate the likelihoods $\mathcal{L}_{i}=\mathcal{L}(X_{i})$, where $X_{i}$ is a
sequence of decreasing values,
\begin{equation}
0<X_{M}<\cdots <X_{2}<X_{1}< X_{0}=1,
\end{equation}
as shown schematically in Fig.~\ref{fig:NS}, the evidence can be approximated numerically using standard quadrature methods as a weighted sum
\begin{equation}
\mathcal{Z}={\textstyle {\displaystyle \sum_{i=1}^{M}}\mathcal{L}_{i}w_{i}},
\label{eq:NS_sum}
\end{equation}
where the weights $w_{i}$ for the simple trapezium rule are given by $w_{i}=\frac{1}{2}(X_{i-1}-X_{i+1})$. An example of a posterior in two dimensions and its associated function $\mathcal{L}(X)$ is shown in
Fig.~\ref{fig:NS}.

The summation in Eq.~(\ref{eq:NS_sum}) is performed as follows. The iteration counter is first set to~$i=0$ and $N$ `active' (or `live') samples are drawn from the full prior $p(\mathbf{P_{MSSM}})$, so the
initial prior volume is $X_{0} = 1$. The samples are then sorted in order of their likelihood and the smallest (with likelihood $\mathcal{L}_{0}$) is removed from the active set (hence becoming `inactive')
and replaced by a point drawn from the prior subject to the constraint that the point has a likelihood $\mathcal{L}>\mathcal{L}_{0}$. The corresponding prior volume contained within the iso-likelihood contour
associated with the new live point will be a random variable given by $X_{1} = t_{1} X_{0}$, where $t_{1}$ follows the distribution $\Pr(t) = Nt^{N-1}$ (i.e., the probability distribution for the largest of
$N$ samples drawn uniformly from the interval $[0,1]$). At each subsequent iteration $i$, the removal of the lowest likelihood point $\mathcal{L}_{i}$ in the active set, the drawing of a replacement with
$\mathcal{L} > \mathcal{L}_{i}$ and the reduction of the corresponding prior volume $X_{i}=t_{i} X_{i-1}$ are repeated, until the entire prior volume has been traversed. The algorithm thus travels through
nested shells of likelihood as the prior volume is reduced. The mean and standard deviation of $\log t$, which dominates the geometrical exploration, are: 
\begin{equation}
E[\log t] = -1/N, \quad \sigma[\log t] = 1/N.
\end{equation}
Since each value of $\log t$ is independent, after $i$ iterations the prior volume will shrink down such that $\log X_{i} \approx -(i\pm\sqrt{i})/N$. Thus, one takes $X_{i} = \exp(-i/N)$. 

Once the evidence~$\mathcal{Z}$ is found, posterior inferences can be easily  generated using the final live points and the full sequence of discarded points from the nested sampling  process, i.e., the
points with the lowest likelihood value at each iteration~$i$ of  the algorithm. Each such point is simply assigned the probability weight
\begin{equation}
p_{i}=\frac{\mathcal{L}_{i}w_{i}}{\mathcal{Z}}.\label{eq:12}
\end{equation}
These samples can then be used to calculate inferences of posterior parameters such as  means, standard deviations, covariances and so on, or to construct marginalised posterior distributions.

The most challenging task in implementing nested sampling is to draw samples from the prior within the hard constraint $\mathcal{L}> \mathcal{L}_i$ at each iteration $i$. The {\sc MultiNest} algorithm
\cite{2008MNRAS.384..449F, 2009MNRAS.398.1601F} tackles this problem through an ellipsoidal rejection sampling scheme. The live point set is enclosed within a set of (possibly overlapping)  ellipsoids and a
new point is then drawn uniformly from the region enclosed by these ellipsoids. The ellipsoidal decomposition of the live point set is chosen to minimize the sum of volumes of the ellipsoids. The ellipsoidal
decomposition is well suited to dealing with posteriors that have curving degeneracies, and allows mode identification in multi-modal posteriors. If there are subsets of the ellipsoid set that do not overlap
with the remaining ellipsoids, these are identified as a distinct mode and subsequently evolved independently. The {\sc MultiNest} algorithm has proven very useful for tackling inference problems in cosmology
and particle physics(see e.g. \cite{2009MNRAS.400.1075B, AbdusSalam:2009qd, 2009PhRvD..80c5017A, 2008JHEP...10..064F, 2008JHEP...12..024T, 2009arXiv0911.1986R, 2009CQGra..26u5003F}) typically showing two
orders of magnitude improvement in efficiency over conventional techniques. We use this algorithm for exploring the posterior distributions in this paper.

\section{Focus point SUSY at the LHC}\label{Results}

\subsection{Conventional input measurements}
Before explaining how to improve the LHC measurements in the focus point region, we will reproduce the existing approach to provide a comparison for our later results. Assuming that the LHC will not produce squarks or sleptons, mass measurements are restricted to cascade decays of the form:
\begin{equation}
\tilde{g}\rightarrow \tilde{\chi}^0_{j} q \bar{q} \rightarrow \tilde{\chi}^0_i q \bar{q} l^+ l^-
\end{equation}
where $j > i$. The neutralino decays give rise to kinematic endpoints in the dilepton invariant mass distribution that will be recognisable as arising from three body decays (by virtue of shape or possibly by using a wedgebox technique as in~\cite{Kang:2009sk}). In fact, two endpoints are visible for the focus point model considered here, arising from decays of the $\tilde{\chi}^0_{2}$ and  $\tilde{\chi}^0_{3}$, and their positions are simply given by the difference $m_{\tilde{\chi}^0_{i}}-m_{\tilde{\chi}^0_{1}}$. The most recent relevant study performed by members of the ATLAS collaboration~\cite{desanctis-2007} gives an error of 0.5 GeV on $m_{\tilde{\chi}^0_{2}}-m_{\tilde{\chi}^0_{1}}$, and 1.5 GeV on $m_{\tilde{\chi}^0_{3}}-m_{\tilde{\chi}^0_{1}}$ if we assume, as these studies have, that we have correctly identified the neutralinos producing the bumps. The resulting distribution is shown in figure~\ref{fig:dilep} and would be very similar for the benchmark point used in this study. We therefore assume for our study that the neutralino mass differences have been determined to be the nominal values obtained from the \tt ISASUSY \rm mass spectrum for our MSSM benchmark point, with the same errors as those quoted in the ATLAS study.
\FIGURE[h!]{\epsfig{file=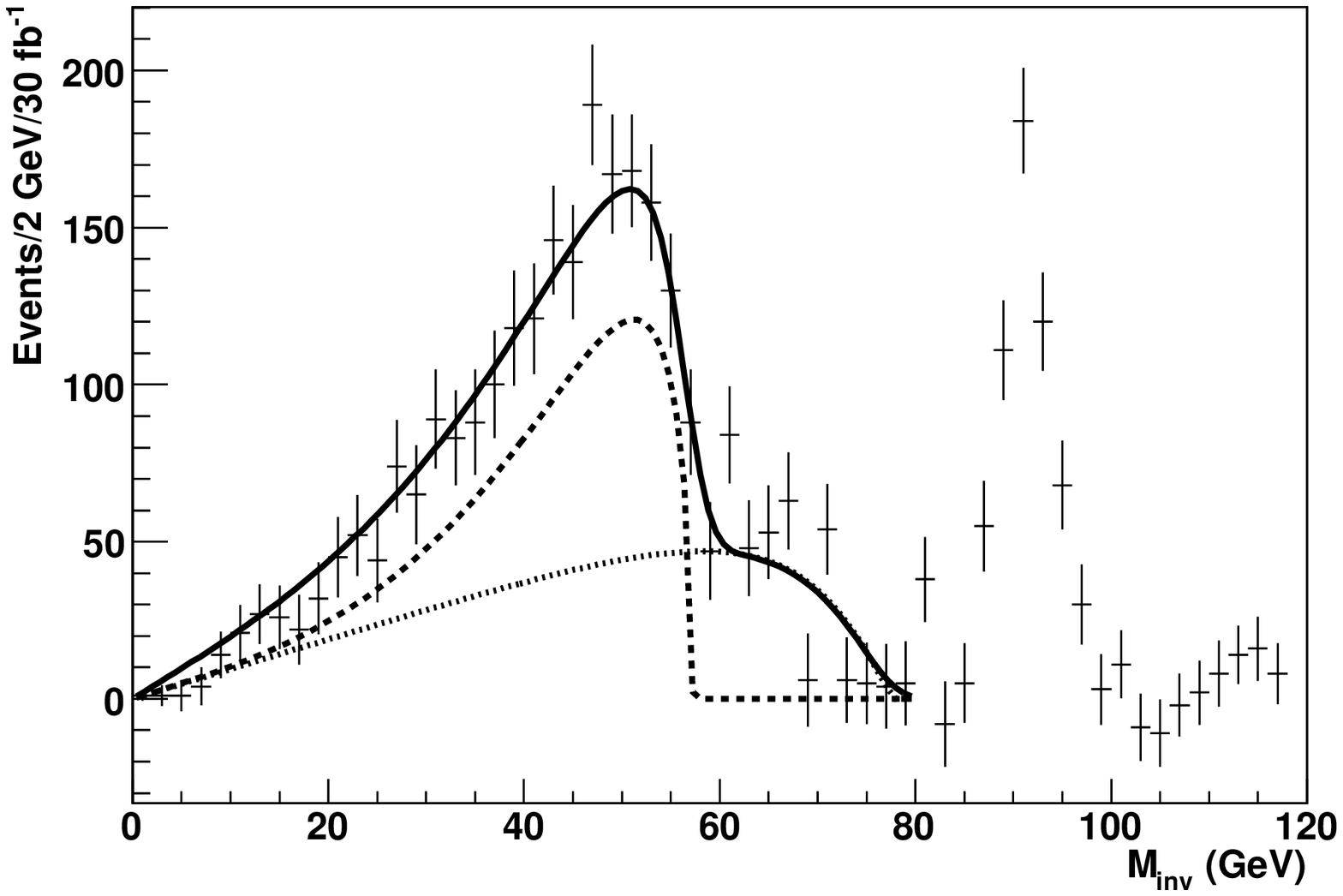,width=4in}\caption{The dilepton\label{fig:dilep} invariant mass reconstructed by the ATLAS detector for a benchmark point in the focus point region (taken from reference~\cite{desanctis-2007}). Two kinematic endpoints are clearly visible and can be measured precisely by fitting with the analytic form of the three body endpoint shape.}}
%\begin{figure}
%\centerline{
%\includegraphics[width=2.6in]{fig10.eps}
%}
%\caption{The dilepton invariant mass reconstructed by the ATLAS detector for a benchmark point in the focus point region (taken from reference~\cite{desanctis-2007}). Two kinematic endpoints are clearly visible and can be measured precisely by fitting with the analytic form of the three body endpoint shape. 
%\label{fig:dilep}}
%\end{figure}

Reference~\cite{baltz-2006-74} further assumes that the $\tilde{\chi}^0_{1}$ and $\tilde{g}$ masses can be fixed to within ~10\% by kinematic fits, and the lightest Higgs mass could be obtained with an accuracy of 0.25 GeV. Again, we assume these same errors, and use the nominal mass values from our benchmark point spectrum for the reconstructed masses. Finally, one can set limits on the squark and slepton masses based on their non-observation, which we impose approximately by restricting the prior range for the squark and slepton mass parameters. A summary of the observables used in our analysis is given in Table~\ref{observables}, while the prior range used in our exploration of the MSSM parameter space is given in Table~\ref{prior-range}.
\clearpage
\TABLE[!h]{
\begin{tabular}{|c|c|}
\hline
Observable&Value\\
\hline
$m_{\tilde{\chi}^0_{2}}-m_{\tilde{\chi}^0_{1}}$&$59.5 \pm 0.5$ GeV\\
$m_{\tilde{\chi}^0_{3}}-m_{\tilde{\chi}^0_{1}}$&$85.3 \pm 1.5$ GeV\\
$m_{\tilde{g}}$&$850 \pm 85$ GeV\\
$m_{h}$&$118.72 \pm 0.25$ GeV\\
$m_{\tilde{\chi_1^0}}$&$106 \pm 11$ GeV\\
\hline
\end{tabular}
\caption{Set of observables \label{observables}used for the `conventional' analysis of our focus point benchmark point. All masses were calcuated using \tt ISASUSY\rm, while the errors are quoted from the studies in~\cite{desanctis-2007} and~\cite{baltz-2006-74}. See text for further explanation.}
}

We have performed a nested sampling of the MSSM parameter space, using a log prior on all parameters except the $A$ parameters for which we follow~\cite{baltz-2006-74} in using the formula $A = M$sinh$(x)$, with $M=50$ GeV. A log prior is generally useful for parameters with a scale uncertainty, while $A = M$sinh$(x)$ is chosen since the $A$ parameters can take either sign. For simplicity, we have further restricted $M_1$ and $\mu$ to be positive as this will not affect the validity of our conclusions. For each point selected by {\sc MultiNest} we run \tt ISASUSY \rm on the 24 MSSM parameters to generate the mass and decay spectrum. We then use the masses and neutralino mass differences to calculate the predicted values of the observables given in Table~\ref{observables} and evaluate the likelihood using equation~\ref{likelihood}.

\TABLE[!h]{
\begin{tabular}{|c|c|}
\hline
Parameters&Prior range\\
\hline
$M_1, M_2, \mu, m_A$&0.1 GeV -- 4000 GeV\\
tan$\beta$&2 -- 60\\
$m_{\tilde{g}}$&700 GeV -- 4000 GeV\\
1st and 2nd generation squark masses&2000 GeV -- 4000 GeV\\
3rd generation squark masses&1500 GeV -- 4000 GeV\\
slepton masses&350 GeV -- 4000 GeV\\
$A_t,A_b,A_c$&-4000 -- 4000\\
\hline
\end{tabular}
\caption{Prior ranges used in all fits for the 24 MSSM parameters.\label{prior-range}}
}

\FIGURE[h!]{\epsfig{file=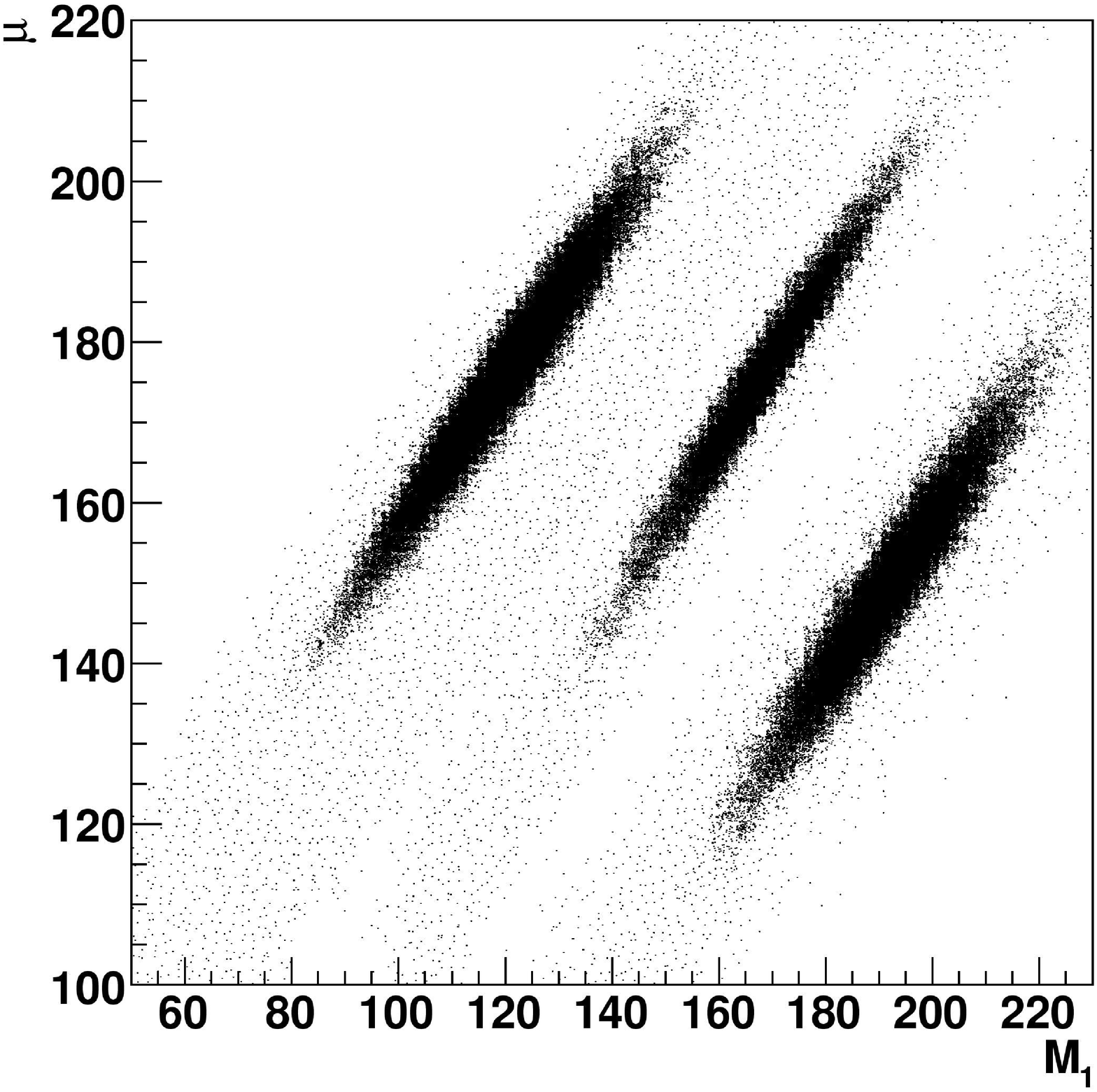,width=4in}\caption{Marginalised posterior in the ($M_1$,$\mu$) plane, resulting from a 24D {\sc MultiNest} sampling of the weak scale MSSM parameters with the `conventional' set of input measurements.\label{m1mu-old}}}

Figure~\ref{m1mu-old} shows the marginalised posterior in the ($M_1$,$\mu$) plane as evaluated by {\sc MultiNest} for this set of input measurements, and one can clearly see three solutions. Each of these has a lightest neutralino of different dominant character, with the left-most island corresponding to the benchmark model\footnote{The results agree closely but not exactly with those shown in~\cite{baltz-2006-74}, but the differences can be explained either by our slightly different procedure of handling the MSSM weak scale spectrum or by the fact that we do not impose a Tevatron-inspired limit on the lightest chargino mass. This latter cut restricted the phase space for the false solutions in~\cite{baltz-2006-74} and, without it, we essentially get extra points in the `wrong' islands which contribute to raising the `wrong peaks' in figures~\ref{relic-old} and~\ref{annihilation-old}.}. One can use the points in this posterior to calculate astrophysical observables, and we used \tt DarkSUSY 5.05\rm~\cite{gondolo-2004-0407,frank-2007-047,degrassi-2003-28,heinemeyer-1998} with a custom interface to \tt ISASUSY \rm based on the Les Houches format~\cite{skands-2004-0407}. The WIMP relic density and neutralino pair annihilation cross-section at threshold (used to calculate the flux expected in indirect search experiments) are shown in Figures~\ref{relic-old} and~\ref{annihilation-old}, and we see that the effect of the extra islands is to give us false regions of high likelihood orders of magnitude away from the correct value. These would clearly destroy any reasonable prospect of accurately predicting the relic density or gamma ray flux based on LHC measurements. Direct search cross-sections divide into spin-dependent and spin-independent contributions, with the spin-independent term enhanced by a factor of the square of the target nucleus mass. The effect of false posterior regions at the LHC is to contribute an extra peak to each of these terms, as shown in Figure~\ref{protonSI-old} for the spin-independent neutralino-proton direct detection cross-section and in Figure~\ref{neutronSD-old} for the spin-dependent neutralino-neutron direct detection cross-section (the spin-dependent proton-neutralino and spin-independent neutralino-neutron cross-sections are very similar). We see a longer tail on the spin dependent distribution than in~\cite{baltz-2006-74}, but this is almost certainly due to {\sc MultiNest} having explored the bino mode more thoroughly than their sampling method.

\DOUBLEFIGURE[h!]{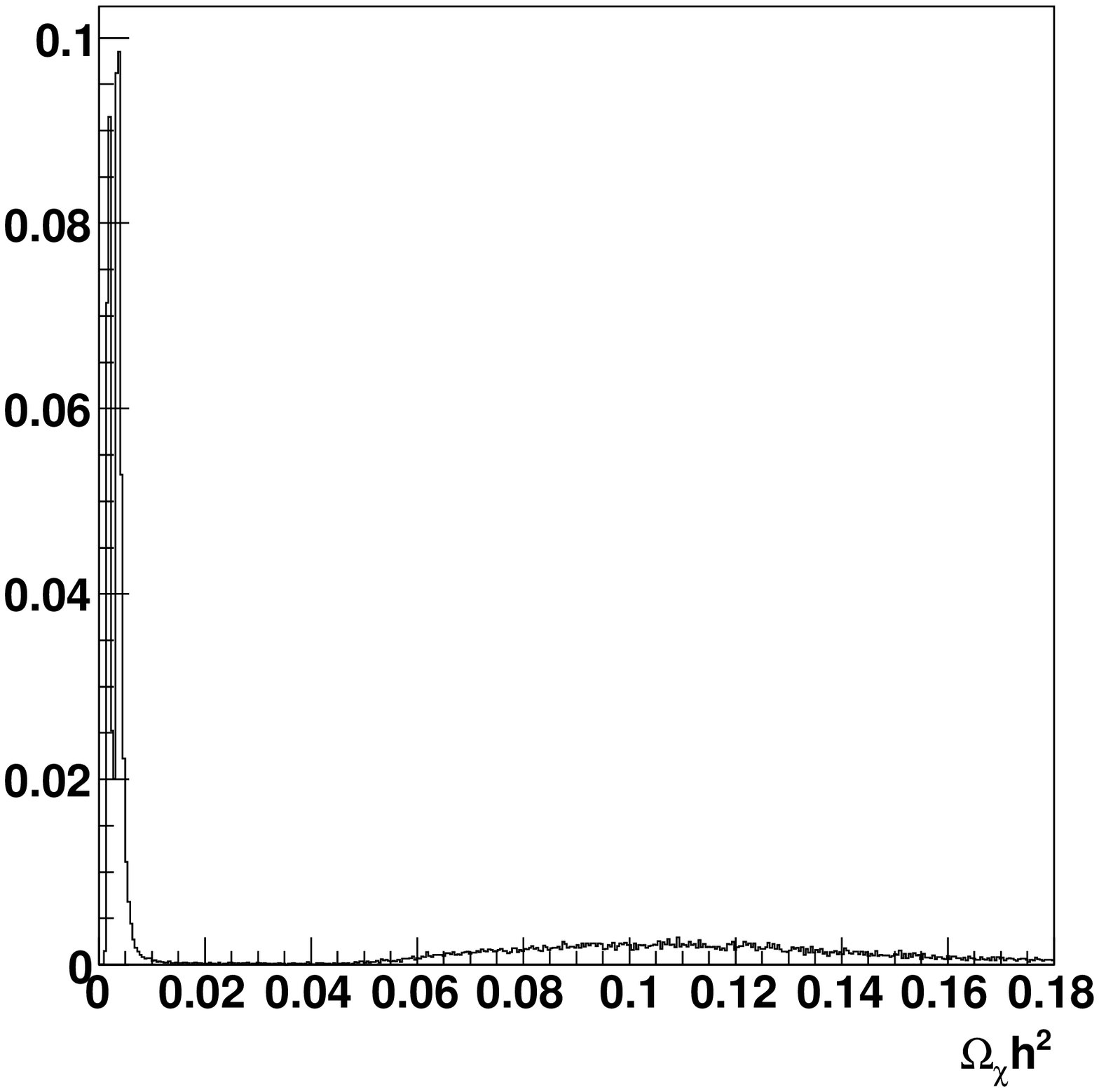,width=2.6in}{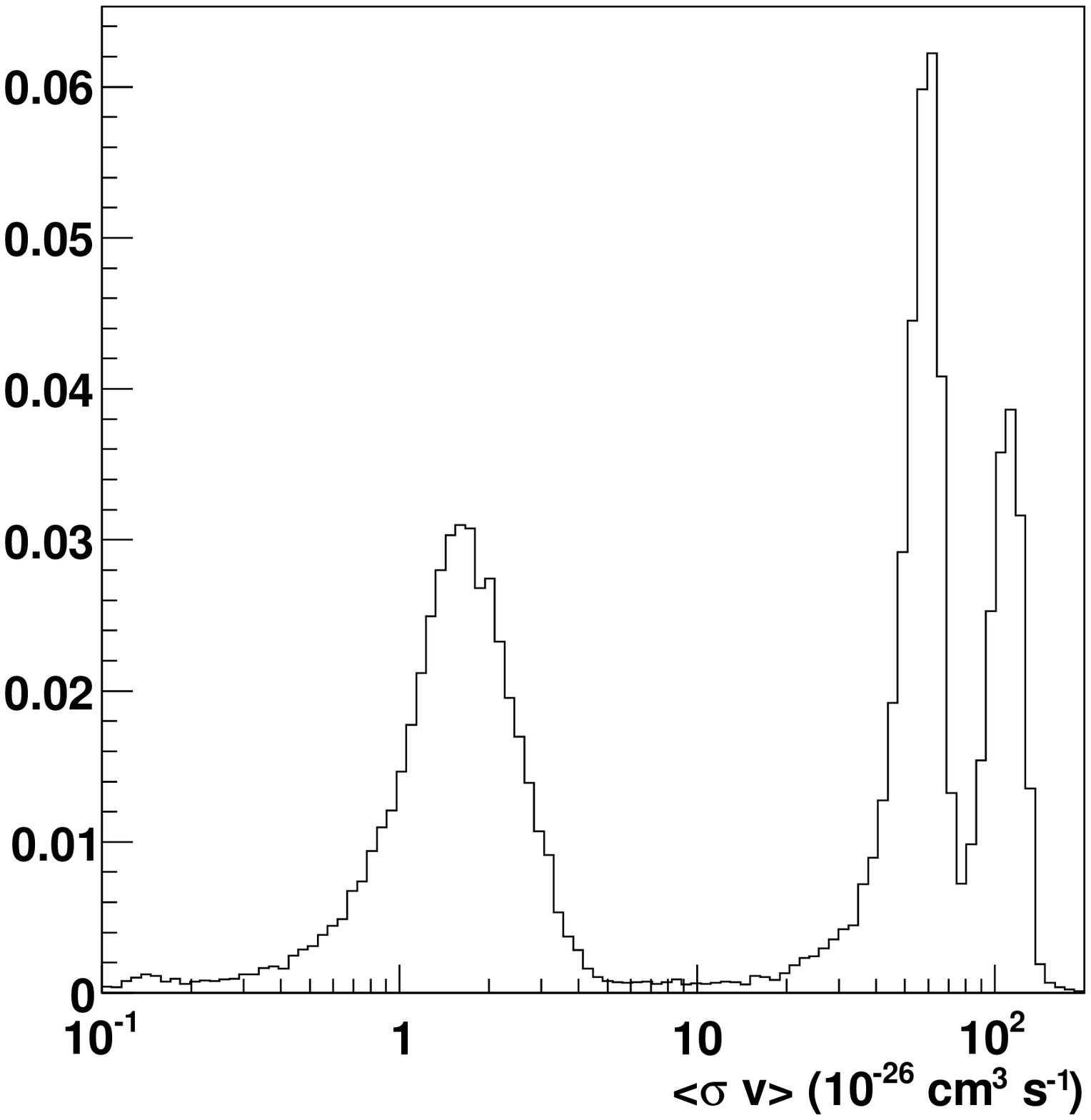,width=2.6in}{The relic density as predicted from the `conventional' set of LHC measurements.\label{relic-old}}{The neutralino pair annihilation cross-section at threshold as predicted from the `conventional' set of LHC measurements.\label{annihilation-old}}
\DOUBLEFIGURE[h!]{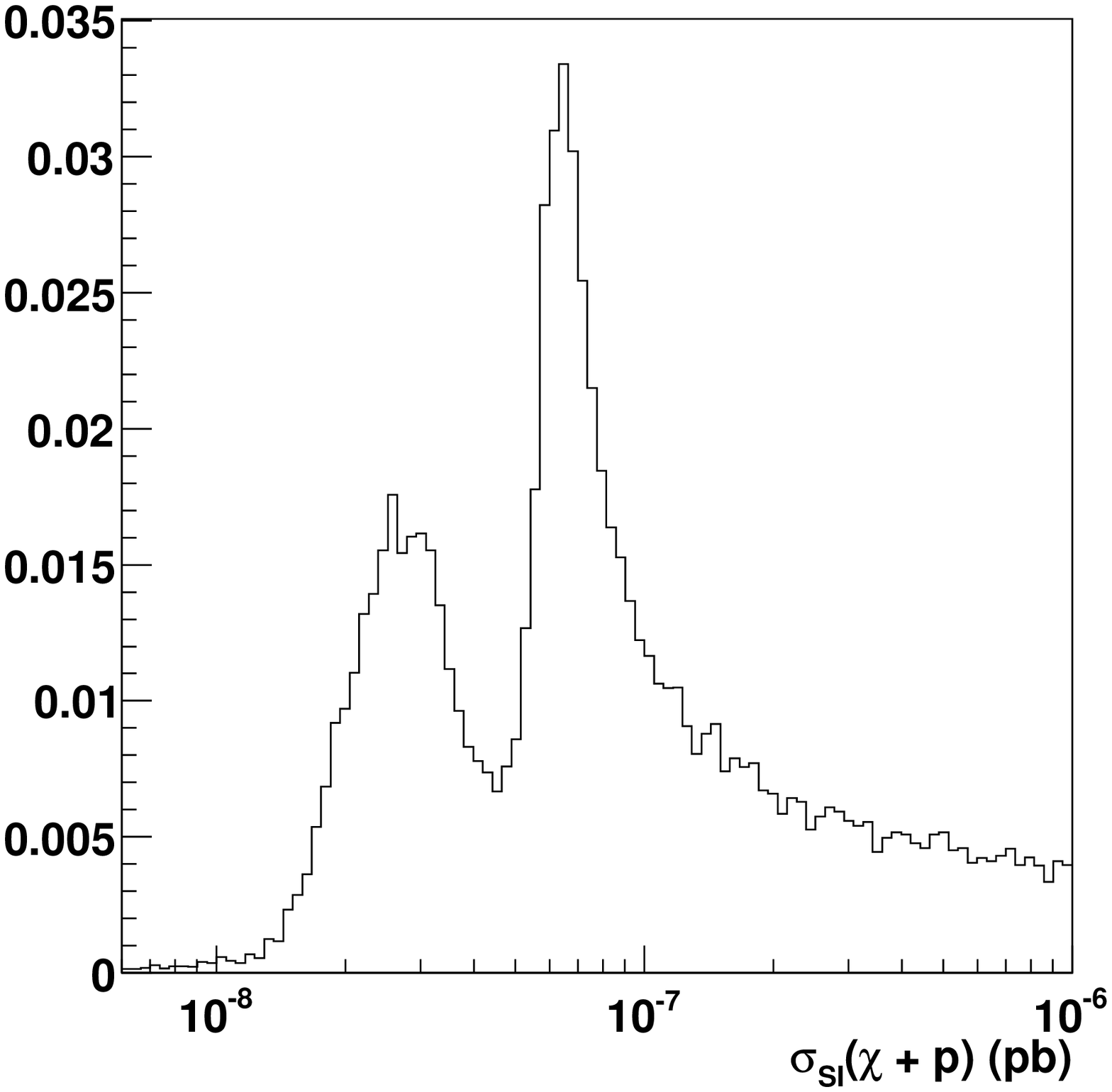,width=2.6in}{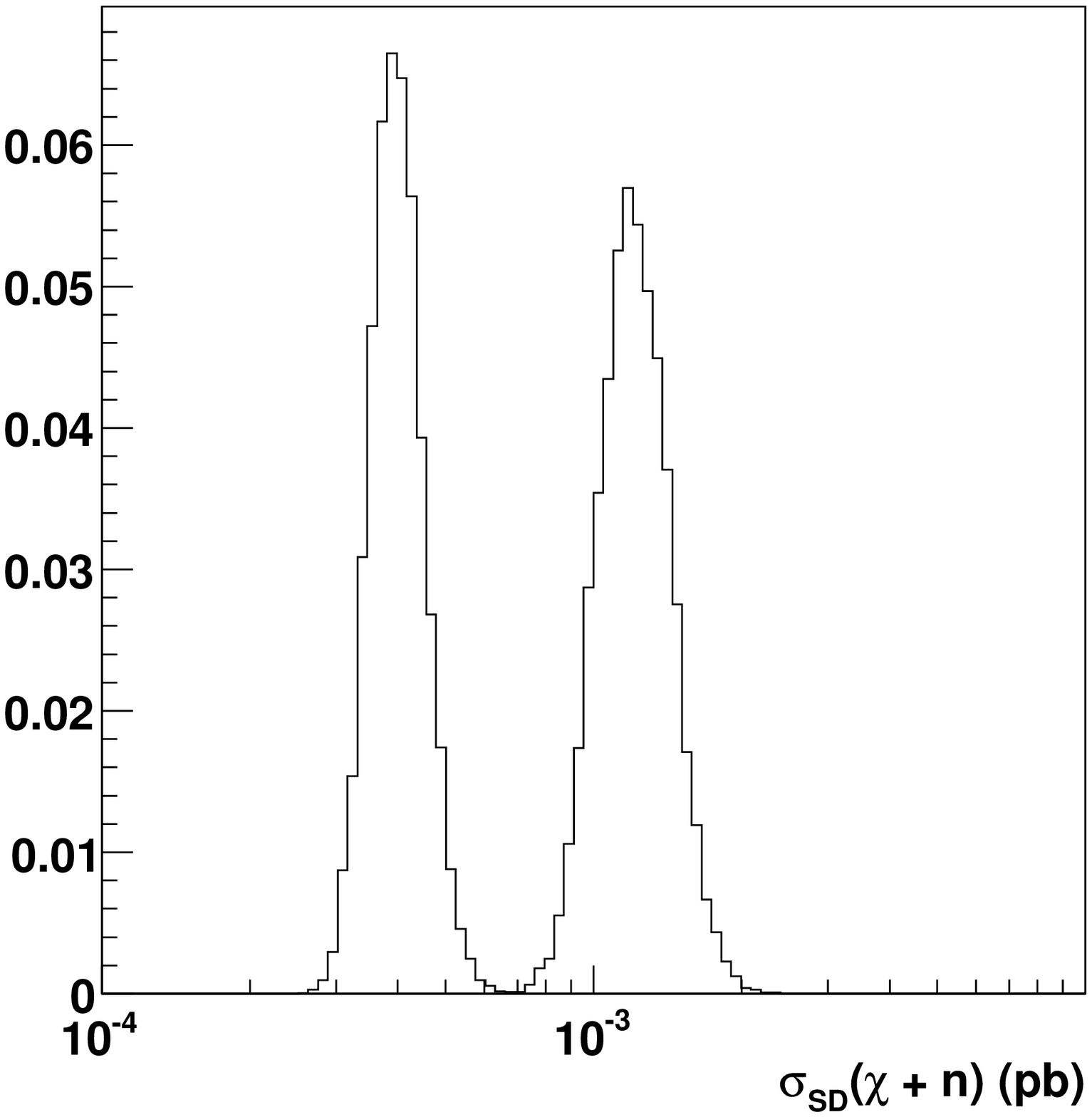,width=2.6in}{The spin-independent neutralino-proton direct detection cross-section as predicted from the `conventional' set of LHC measurements.\label{protonSI-old}}{The spin-dependent neutralino-neutron direct detection cross-section as predicted from the `conventional' set of LHC measurements.\label{neutronSD-old}}
\subsection{New input measurements}
To remove the false regions of the posterior in Figure~\ref{m1mu-old}, we need to find observables at the LHC that are sensitive to neutralino mixing. Here we suggest one example in the form of the shape of the dilepton invariant mass distribution, since the number of events contributing to each bump in the distribution is sensitive to the couplings of the relevant neutralinos producing the bumps. Tuning the WIMP components has the effect of raising and lowering the two bumps, and thus changing the shape. In general, tuning the neutralino couplings could also add or remove endpoints, or change their interpretation (i.e. making an endpoint appear from a $\tilde{\chi}^0_{4} \rightarrow \tilde{\chi}^0_1l^+l^-$ decay rather than from a $\tilde{\chi}^0_{3} \rightarrow \tilde{\chi}^0_1l^+l^-$, etc). Although the dependence of the shape on the neutralino components is non-trivial, we are helped in the case of heavy scalars by the fact that we know from the LHC data that there are no on shell sleptons contributing to the endpoints, and thus the shape is even more sensitive to the neutralino couplings than in the generic case where one would complicate matters by introducing greater dependence on the slepton masses.

In principle, one could model the shape of the dilepton invariant mass precisely at a given point in parameter space by generating a large sample of Monte Carlo events at that point, and passing them through a detector simulation. For a thorough sampling of the 24 parameter MSSM involving millions of explored points, this is clearly unfeasible, however, and we must use some shortcuts to encode the shape information in a manner that is easy to calculate. Under the assumption that the three body neutralino decays proceed dominantly through an off-shell $Z$ boson~\footnote{Our decay through $Z$ boson assumption is in fact true in the focus point region since the sleptons are very heavy, though \emph{a priori} one can only set lower bounds on the slepton masses and hence might encounter a systematic uncertainty due to this asssumption. In practise, however, we shall find the ratio of the number of events contributing to each bump in the invariant mass distribution is $\approx$50 times higher in one of the false solutions, and $\approx$5 times higher in the other which probably exceeds any systematic correction to the ratio measurement.}, the shape of the dilepton invariant mass distribution is a known function of the neutralino masses, whose fit to the measured distribution is shown in Figure~\ref{fig:dilep}. The integral can be used to calculate the ratio of the number of events arising from each neutralino decay contributing to the dilepton invariant mass, which is related to the branching ratios of the sparticles by:
\begin{equation}
\frac{BR(\tilde{g} \rightarrow ... \rightarrow \tilde{\chi}^0_{2} + X) \times BR(\tilde{\chi}^0_{2} \rightarrow \tilde{\chi}^0_1 l^+ l^-)}{BR(\tilde{g} \rightarrow ... \rightarrow \tilde{\chi}^0_{3} + X) \times BR(\tilde{\chi}^0_{3} \rightarrow \tilde{\chi}^0_1 l^+ l^-)}
\label{ratioequation}
\end{equation}
where the first decay in the numerator and denominator can proceed through multiples steps provided they do not produce leptons. The ATLAS study in reference~\cite{desanctis-2007} quotes a measured value of $1.4 \pm 0.3$ for this ratio, comparing favourably to their theoretical value of 1.19, even in the presence of background and a detailed simulation of the ATLAS detector resolution. Our benchmark point has a slightly different theoretical value of 1.46. The study in~\cite{desanctis-2007} further assumed that the $\tilde{\chi}^0_{3}$ and $\tilde{\chi}^0_{2}$ had been correctly identified as the cause of the endpoints in the dilepton invariant mass spectrum, an assumption that we do not wish to work with here since it presupposes knowledge of neutralino mixing which we are trying to determine in a general manner. 

There are other more trivial details of the shape of the dilepton invariant mass distribution. It is clear that any parameter point capable of explaining the distribution in figure~\ref{fig:dilep} must have a taller endpoint on the left, and a shorter endpoint on the right (this is of course the same as using ratio information). It is also clear that \emph{only} two endpoints were observed, and hence any model that would have given more or less than two endpoints is also not capable of explaining the observed distribution. 

If we now assume we have not identified the neutralinos producing the endpoints, we must examine the general case where the number of $\tilde{\chi}^0_{j} \rightarrow \tilde{\chi}^0_i l^+ l^-$ decays contributing to the dilepton invariant mass distribution is given by:
\begin{equation}
N \propto BR(\tilde{g} \rightarrow ... \rightarrow \tilde{\chi}^0_{j} + X) \times BR(\tilde{\chi}^0_{j} \rightarrow \tilde{\chi}^0_i l^+ l^-)
\label{BRequation}
\end{equation}
where the exact number of events depends on the cross-section for gluino production and the efficiency of the kinematic cuts. Again, the first branching ratio shown could in fact be the product of various decay processes that end with a $\tilde{\chi}^0_{j}$ and anything else (excluding leptons), and could consist of multiple steps. Since we know (from measurement) that all decay chains contributing to our dilepton plot started with gluinos, however, we can confidently state that we would have seen two endpoints if the two largest combined branching ratios of the form of equation~\ref{BRequation} are close in value while the third highest branching ratio is much lower. This encodes the shape of the distribution in a way which is easy to evaluate quickly at many points in parameter space, making it possible to scan or sample the parameter space to reject points on the basis of shape.

We therefore reran the {\sc MultiNest} sampling with the following additions. At each point in the parameter space, we used the \tt ISASUSY \rm decay spectrum to work out the combined branching ratios for all processes of the form given in equation~\ref{BRequation}. The two highest of these branching ratios are assumed to have caused the observed bumps in the dilepton invariant mass distribution, and the ratio of the number of events contributing to each endpoint is obtained using the following generalisation of equation~\ref{ratioequation}: 

\begin{equation}
\frac{BR(\tilde{g} \rightarrow ... \rightarrow \tilde{\chi}^0_{p} + X) \times BR(\tilde{\chi}^0_{p} \rightarrow \tilde{\chi}^0_q l^+ l^-)}{BR(\tilde{g} \rightarrow ... \rightarrow \tilde{\chi}^0_{r} + X) \times BR(\tilde{\chi}^0_{r} \rightarrow \tilde{\chi}^0_s l^+ l^-)}
\end{equation}

The gluino cross-section and reconstruction efficiency cancel in this division and thus do not need to be explicitly evaluated- we are indeed fortunate that squarks are not contributing otherwise this procedure would not have worked. Note that our two highest branching ratios can in principle come from any neutralino decay process, so in principle we have endpoints arising from the decays $\tilde{\chi}^0_{p} \rightarrow \tilde{\chi}^0_q l^+ l^-$ and $\tilde{\chi}^0_{r} \rightarrow \tilde{\chi}^0_s l^+ l^-$, where $p$,$q$,$r$ and $s$ can take any value as long as $p>q$ and $r>s$ (in the previous section, we had $p=2$, $r=3$ and $q=s=1$). We now assign a likelihood based on the observables listed in Table~\ref{newobservables}, which simply comprise of generalisations of those in the previous section plus the addition of the ratio measurement. Furthermore, we assign zero likelihood to points that could not have produced the observed dilepton distribution because they fell into the following categories:
\begin{enumerate}
\item Points in which the `wrong' endpoint is highest, i.e. the kinematic endpoint with the largest $m_{ll}$ value is taller than the endpoint with the lowest $m_{ll}$ value\\
\item Points in which the third highest combined branching ratio of the form of equation~\ref{BRequation} is more than 10\% of the second highest branching ratio (this is considered a sensible boundary of visibility). \\
\end {enumerate}
\TABLE[h!]{
\begin{tabular}{|c|c|}
\hline
Observable&Value\\
\hline
$m_{\tilde{\chi}^0_{p}}-m_{\tilde{\chi}^0_{q}}$&$59.5 \pm 0.5$ GeV\\
$m_{\tilde{\chi}^0_{r}}-m_{\tilde{\chi}^0_{s}}$&$85.3 \pm 1.5$ GeV\\
$m_{\tilde{g}}$&$850 \pm 85$ GeV\\
$m_{h}$&$118.72 \pm 0.25$ GeV\\
$m_{\tilde{\chi_1^0}}$&$106 \pm 11$ GeV\\
$N_p/N_r$&$1.5 \pm 0.3$\\
\hline
\end{tabular}
\caption{Set of observables \label{newobservables}used for the `new' analysis of our focus point benchmark point. All masses were calcuated using \tt ISASUSY\rm, while the errors are quoted from the studies in~\cite{desanctis-2007} and~\cite{baltz-2006-74}. $N_p/N_r$ is the ratio of the number of events arising from the decays $\tilde{\chi}^0_{p} \rightarrow \tilde{\chi}^0_r l^+ l^-$ and $\tilde{\chi}^0_{r} \rightarrow \tilde{\chi}^0_s l^+ l^-$, measured by fitting a functional form to the dilepton invariant mass distribution. See text for further explanation.}
}
Results of a {\sc MultiNest} sampling in the same parameter space as the previous sub-section are shown in the ($M_1$,$\mu$) plane in figure~\ref{m1mu-new}, and the extra regions have been completely removed. Further investigation reveals that the vetos listed above are not the dominant contribution to this improvement, since both leave many points in both of the incorrect regions. Rather, it is the ratio measurement that provides enough discrimination between them to improve the results, leading to a substantial decrease in likelihood in the false regions. Since this is the ratio between the two highest endpoints produced by the gluino cascade processes, this is really telling us that the ``two bump'' feature of the dilepton invariant mass plot is a feature of the focus point model. The large spread in this ratio between the different regions also indicates that even a reasonably imprecise measurement of this quantity would be enough to favour the correct region. We also note that the selected region has endpoints arising from $\tilde{\chi}^0_{2} \rightarrow \tilde{\chi}^0_1 l^+ l^-$ and $\tilde{\chi}^0_{3} \rightarrow \tilde{\chi}^0_1 l^+ l^-$ decays and hence, although we had not assumed this advance, we have successfully inferred it from the data.

The relic density and annihilation cross-section are shown in figures~\ref{relic-new} and~\ref{annihilation-new}, where it can immediately be seen that the `fake' solution peaks are removed from the annihilation cross-section prediction, thus allowing an accurate comparison of LHC data with the forthcoming round of gamma ray results. The relic density prediction is improved but still suffers from a lack of constraint at low values which is caused by the effect of the $A$ pole; resonant annihilation through the $A$ boson leads to a large reduction of the relic density for any model in which $m_A \approx 2m_{\tilde{\chi}_1^0}$. The direct search cross-sections, meanwhile, enjoy a similar improvement to the annihilation cross-section, as seen in Figures~\ref{protonSI-new} and~\ref{neutronSD-new}, though the spin-dependent cross-section has a tail to large values in which it is possible to get a larger cross-section by increasing the amount of bino-Higgsino mixing or by increasing tan$\beta$~\footnote{We thank Michael Peskin for this suggestion.}. This would seem to be irreducible at the LHC. 

\FIGURE[h!]{\epsfig{file=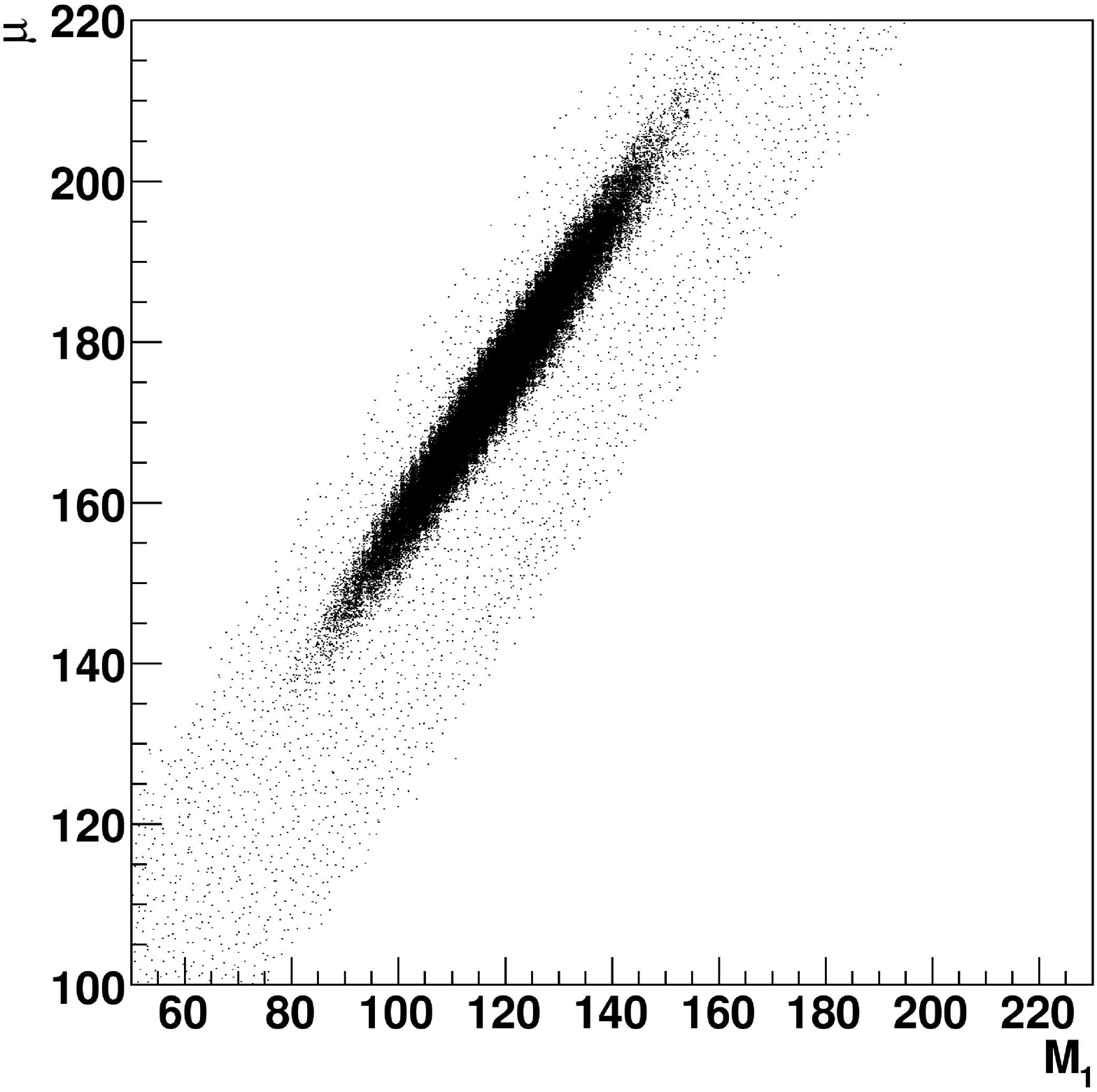,width=4in}\caption{Marginalised posterior in the ($M_1$,$\mu$) plane, resulting from a 24D {\sc MultiNest} sampling of the weak scale MSSM parameters with the `new' set of input measurements.\label{m1mu-new}}}

\DOUBLEFIGURE[h!]{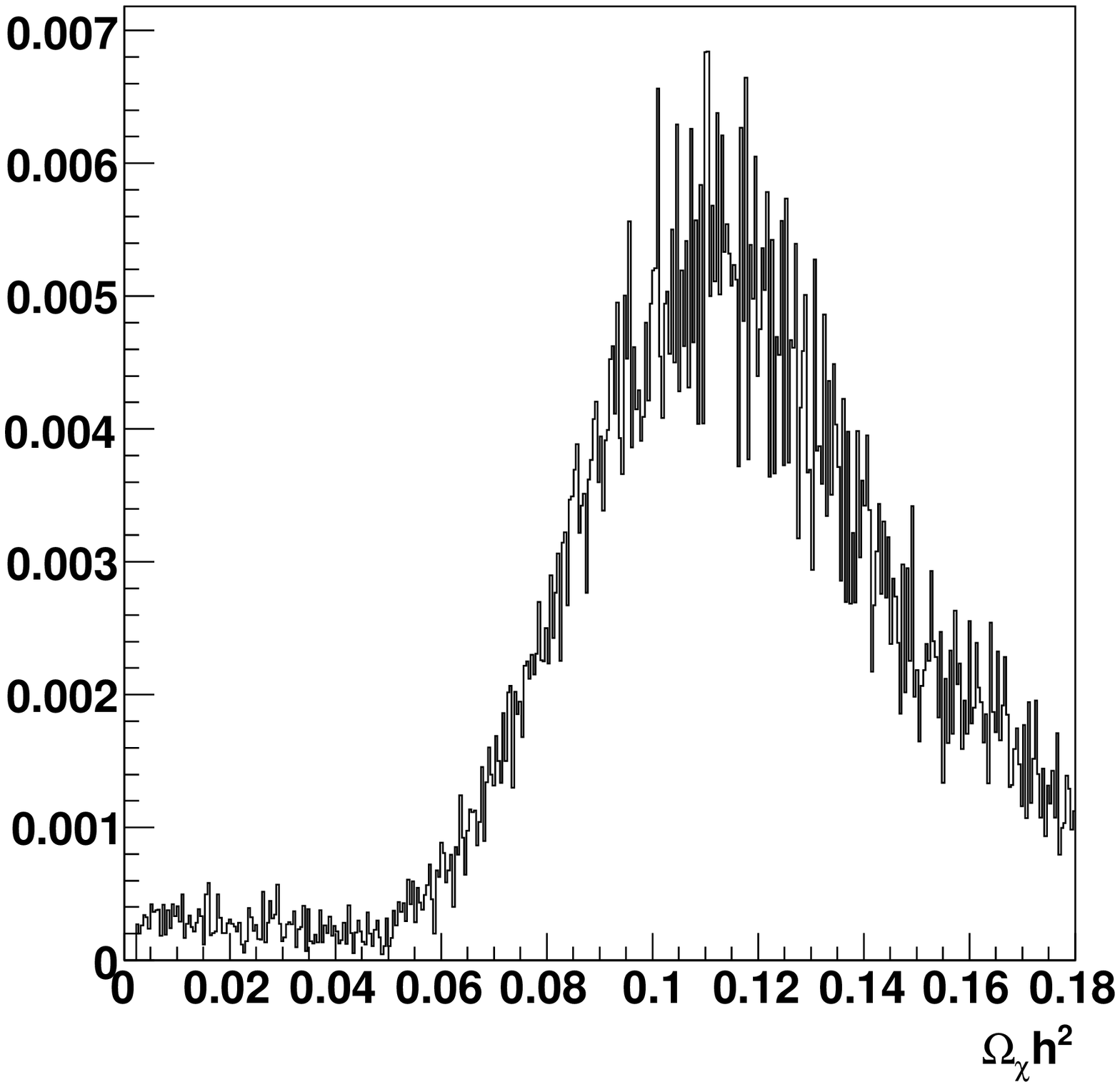,width=2.6in}{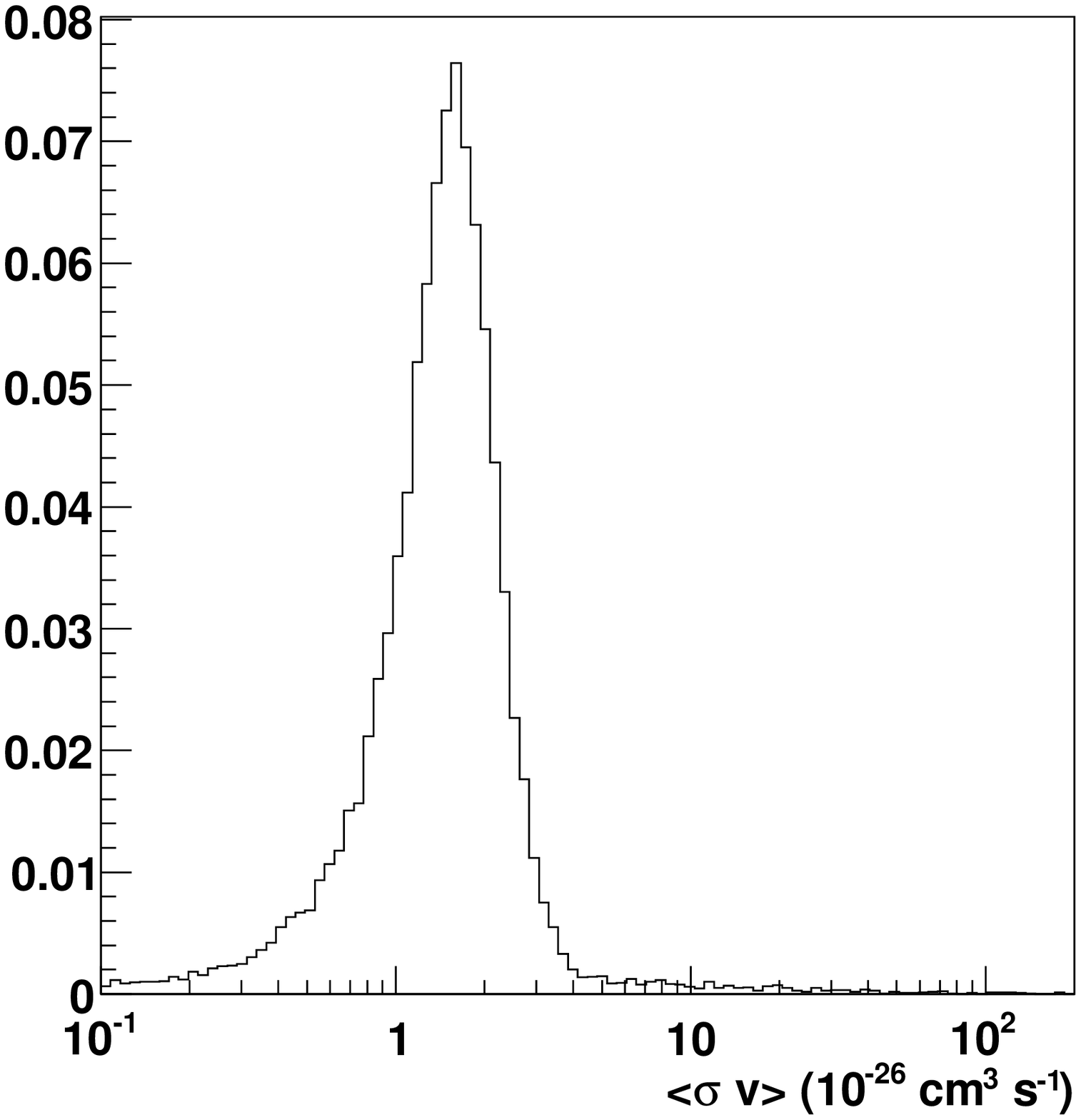,width=2.6in}{The relic density as predicted from the `new' set of LHC measurements.\label{relic-new}}{The neutralino pair annihilation cross-section at threshold (right) as predicted from the `new' set of LHC measurements.\label{annihilation-new}}

\DOUBLEFIGURE[h!]{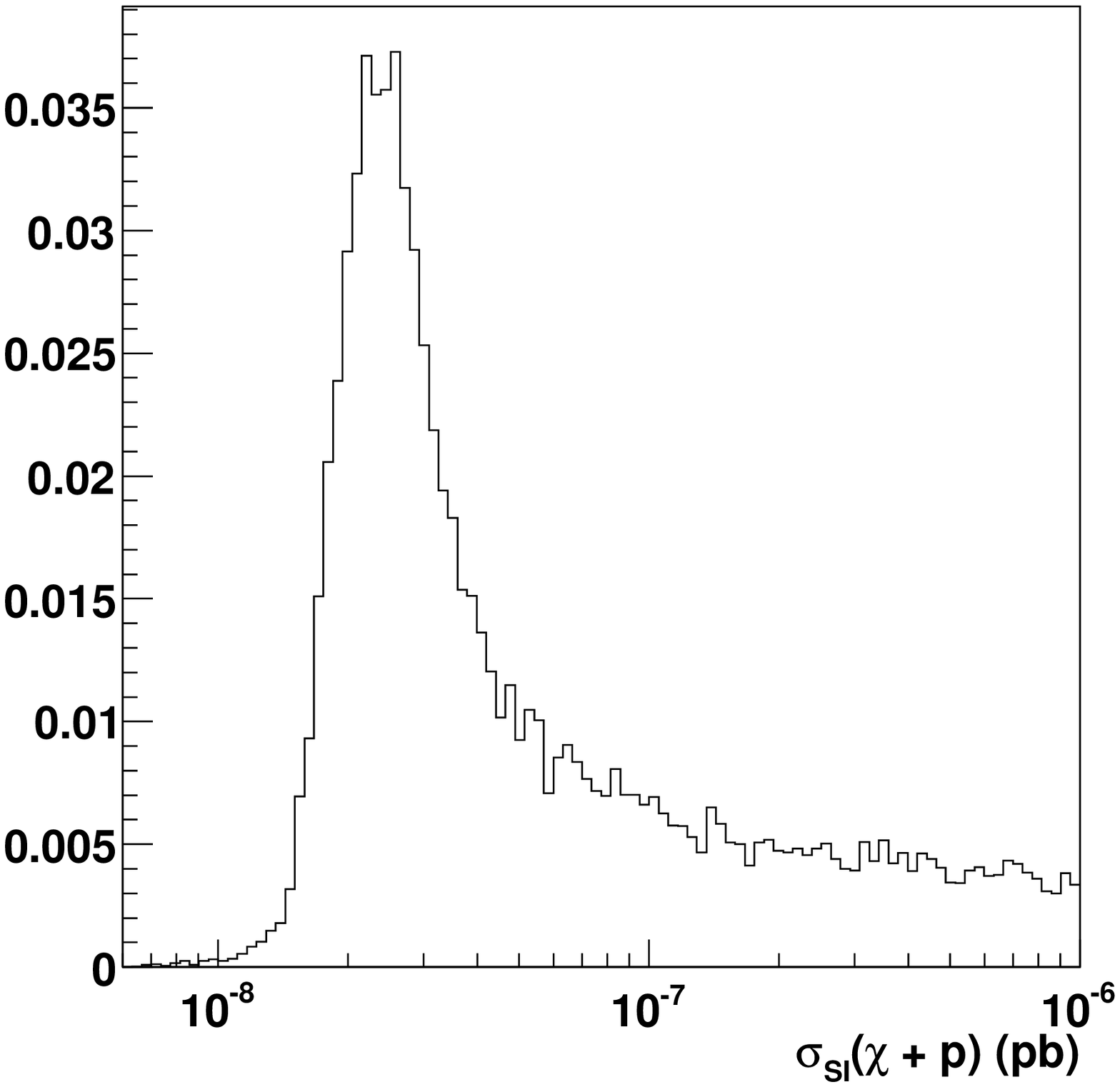,width=2.6in}{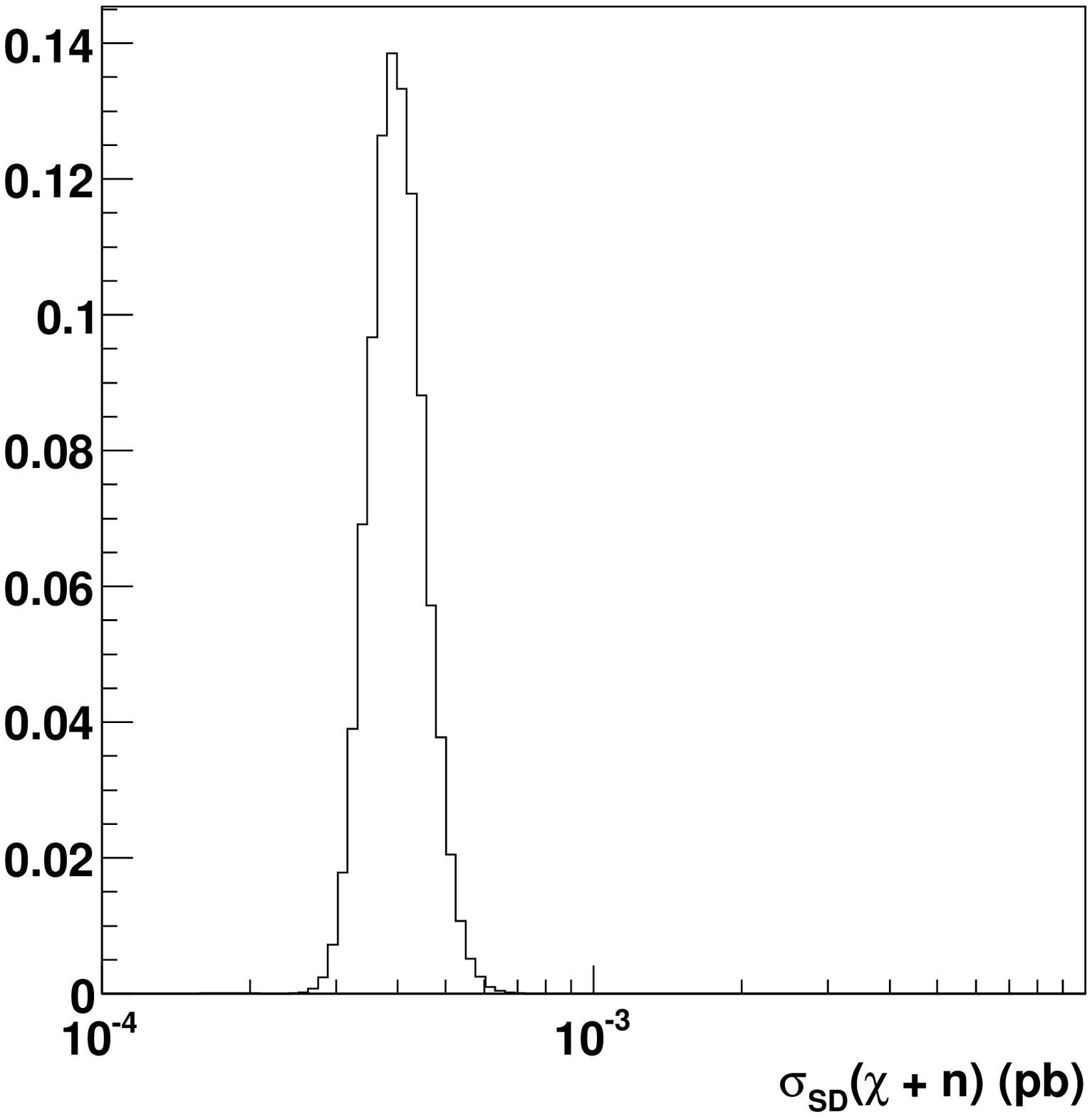,width=2.6in}{The neutralino-proton direct detection cross-section as predicted from the `conventional' set of LHC measurements.\label{protonSI-new}}{The neutralino-neutron direct detection cross-section as predicted from the `conventional' set of LHC measurements.\label{neutronSD-new}}

\subsection{Discussion}\label{Discussion}
Having demonstrated an improvement in the LHC capability at one benchmark point, we now consider the generality of our result, along with any potential pitfalls. Firstly, it is clear that the improvement in our results comes from using shape information, and one must therefore carefully consider whether there are issues that might distort this shape. The dilepton invariant mass is a particularly clean signature, since practically all SUSY backgrounds can be removed via `flavour-subtraction' (i.e. one plots the combination $e^{+}e^{-} + \mu^{+}\mu^{-} - e^{+}\mu^{-} + \mu^{+} e^{-}$). This explains the precision of the quoted ATLAS result which survived a rigorous detector simulation. 

It is more likely that the method described here will fail if Nature presents a focus point scenario with larger mass differences between the neutralinos such that the endpoints are either obscured by the Z peak or, in the case where the mass difference exceeds the Z mass, are not produced at all since the decay $\tilde{\chi}^0_{j}\rightarrow Z \tilde{\chi}^0_{1}$ is open. One could also push the gluino mass to large values and reduce the overall production cross-section at the LHC without violating any current constraints. In such a case, the LHC is trivially bound to fail.

On the plus side, it is possible that the precision of the measurement could be improved further by using inclusive signatures in addition to the endpoint information, using a method similar to that in reference~\cite{lester-2006-0601}. The focus point is conceptually simpler than most other regions of the parameter space, as the fact that the LHC would only be producing gauginos means that any inclusive signature must be telling us something about the gaugino sector. Thus, if dark matter is a gaugino, the focus point could be one of the more promising cases to handle at the LHC rather than one of the hardest, and this is a promising avenue for future work.

Finally, we consider the relic density prediction, and speculate as to whether the low values arising from the $A$ pole can be removed by any further LHC data. Although $m_A$ is well in excess of $2m_{\tilde{\chi_1^0}}$ in the focus point region, there is no way \emph{a priori} to rule out low heavy Higgs masses. If one could measure a lower bound on $m_A$, it is worth noting that the low values are substantially reduced for $m_A> 300$~GeV (Figure~\ref{relic-300}).

\FIGURE[h!]{\epsfig{file=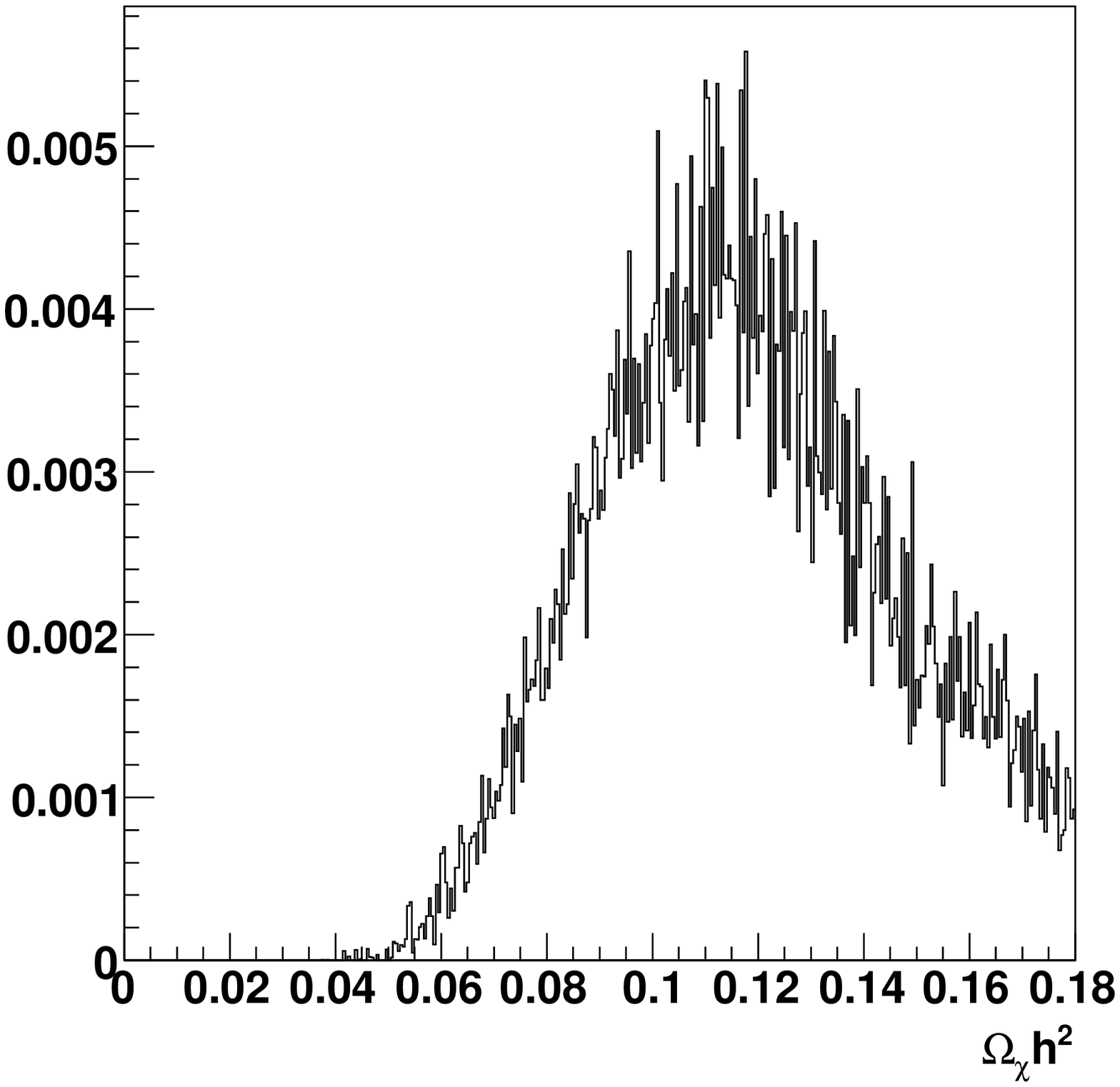,width=2.6in}\caption{The relic density as predicted from the `new' set of LHC measurements with a cut $m_A>300$~GeV.\label{relic-300}}}

Previous studies have placed bounds on the Higgs mass by either examining the Higgs search reach for the LHC in the $m_{A}-$tan$\beta$ plane and determining which region is compatible with non-observation or by using the measured mass difference between sparticles to place bounds on the Higgs mass based on its non-appearance in cascade decays~\cite{nojiri-2006-0603}. Neither method is particularly useful here since a rather large region of the $m_{A}-$tan$\beta$ search plane is compatible with the observation of only $m_h$, while $(m_{\tilde{\chi_4^0}}-m_{\tilde{\chi_1^0}})<2m_{\tilde{\chi_1^0}}$ for our benchmark point, meaning that even if the branching ratios were favourable enough to produce heavy Higgs bosons in cascade decays, the only lower limit we could place on the $A$ mass would still be insufficient to rule out the low values. We therefore consider it unlikely that the LHC could set a lower limit sufficient to reduce this uncertainty on the relic density calculation, though it is not inconceivable that new work on Higgs decays to SUSY particles might extend the search reach for the $A$ boson sufficiently to improve general limits. If evidence suggests that a focus point scenario is realised in Nature, there is a clear incentive to pursue this line of enquiry. 

\section{Conclusions}\label{Conclusions}
We have revisited the case of focus point neutralino dark matter at the LHC, for which it was previously assumed that it was not possible to make accurate predictions of astrophysical observables due to poorly constrained neutralino mixing parameters. Using similar Bayesian sampling techniques as the previous literature, we have demonstrated a method by which the shape of the dilepton invariant mass distribution can be used to remove false solutions and substantially improve predictions of the annihilation cross-section at threshold and the direct search cross-sections. The relic density prediction is also improved and could be significantly enhanced if one could set a lower bound on the $A$ mass.  If Nature has chosen a focus point scenario, we will thus be able to test the compatibility of LHC WIMP measurements with astrophysical data without having to wait for a linear collider. This is particularly important given the rich astrophysical data sets expected over the next few years, and given the high likelihood of obtaining a strong indirect detection signal in focus point models. 

Although there are still specific reasons why we might be unlucky at the LHC (we are still dependent on favourable details of the mass spectrum), we remain optimistic that the LHC will be able to make stronger statements on neutralino mixing than has previously been assumed if Nature has chosen a model with heavy scalars. Applying our technique to other such scenarios would be an interesting line of future work.
\section{Acknowledgements}
MJW gratefully acknowledges the financial support of the Science and Technology Facilities Council, and wishes to thank Howie Baer, Michael Peskin, Edward Baltz, Are Raklev, Giacomo Polesello, Csaba Balazs and the members of the Cambridge Supersymmetry Working group for helpful discussions. FF is supported by a research fellowship from Trinity Hall, Cambridge.
\bibliography{DM_jhep}
\end{document}